\documentclass[11pt]{article}

\usepackage[T1]{fontenc}
\usepackage[utf8]{inputenc}
\usepackage[english]{babel}
\usepackage{algorithm}
\usepackage{algpseudocode}
\usepackage{doi}

\usepackage{tabularx}
\usepackage{array}
\newcolumntype{Y}{>{\raggedright\arraybackslash}X}

\usepackage{booktabs}
\usepackage{longtable}
\usepackage{amssymb} 
\usepackage{amsthm}

\newtheorem{proposition}{Proposition}

\usepackage{tikz}
\usetikzlibrary{positioning, shapes, arrows.meta}

\usepackage[a4paper,margin=30mm]{geometry}
\usepackage{lmodern}
\usepackage{microtype}
\usepackage{setspace}
\usepackage{amsmath,amssymb}

\usepackage{booktabs}
\usepackage{tabularx}

\usepackage[numbers]{natbib}
\usepackage{placeins} 

\title{
	Feasibility-First Satellite Integration in Robust Portfolio Architectures
}

\author{
	Roberto Garrone\\[0.5ex]
	\small
	University of Salford, Salford, UK\\
	\small
	University Sapienza Unitelma, Rome, Italy\\
	\small
	University of Milan-Bicocca, Milan, Italy\\
	\small
	\texttt{roberto.garrone@unimib.it}
}

\date{\today}

\begin{document}
	\maketitle
	
\begin{abstract}
The integration of thematic satellite allocations into core-satellite portfolio architectures is commonly approached using factor exposures, discretionary convictions, or backtested performance, with feasibility assessed primarily through liquidity screens or market-impact considerations. While such approaches may be appropriate at institutional scale, they are ill-suited to small portfolios and robustness-oriented allocation frameworks, where dominant constraints arise not from return predictability or trading capacity, but from fixed costs, irreversibility risk, and governance complexity.

This paper develops a feasibility-first, non-predictive framework for satellite integration that is explicitly scale-aware. We formalize four nested feasibility layers (physical, economic, structural, and epistemic) that jointly determine whether a satellite allocation is admissible. Physical feasibility ensures implementability under concave market-impact laws; economic feasibility suppresses noise-dominated reallocations via cost-dominance threshold constraints; structural feasibility bounds satellite size through an explicit optionality budget defined by tolerable loss under thesis failure; and epistemic feasibility limits satellite breadth and dispersion through an entropy-based complexity budget.

Within this hierarchy, structural optionality is identified as the primary design principle for thematic satellites, with the remaining layers acting as robustness lenses rather than optimization criteria. The framework yields closed-form feasibility bounds on satellite size, turnover, and breadth without reliance on return forecasts, factor premia, or backtested performance, providing a disciplined basis for integrating thematic satellites into small, robustness-oriented portfolios.
\end{abstract}

\noindent\textbf{Keywords:}
core–satellite portfolios; thematic investing; feasibility constraints; structural optionality; robustness-oriented allocation; market impact; portfolio governance; entropy-based complexity; small portfolios; non-predictive frameworks.

\section{Introduction}

The increasing popularity of thematic investing—particularly in areas such as artificial intelligence, digital infrastructure, and technological transformation—has led to widespread adoption of core--satellite portfolio architectures \citep{brinson1986determinants,ilmankirton2012}. In these frameworks, a diversified core is complemented by one or more thematic satellites intended to capture long-horizon structural trends. While this design is intuitively appealing, the integration of satellites remains weakly formalized in the academic literature, especially for small portfolios and robustness-oriented allocation schemes\citep{michaud1989markowitz,deMiguel2009}.

Most existing approaches implicitly treat satellites as scaled-down versions of active strategies: weights are selected using factor exposures, thematic classifications, or discretionary convictions, and feasibility is assessed primarily through liquidity screens or backtested performance. This treatment conflates three conceptually distinct questions:
(i) whether a thematic exposure is economically meaningful,
(ii) whether it can be implemented robustly at a given portfolio scale, and
(iii) whether its integration preserves the governance and interpretability\citep{gennaioli2018crash} of the overall portfolio.

This paper argues that, for small core--satellite portfolios, the dominant constraints on satellite integration are not return predictability, factor structure, or even market impact, but rather feasibility constraints operating at multiple layers: physical, economic, balance-sheet, and epistemic. When portfolio scale is modest, standard market-microstructure constraints are typically slack, while fixed costs, irreversibility risk, and complexity dominate. As a result, optimization-based or signal-driven satellite constructions often introduce fragility without commensurate benefits.

We therefore propose a feasibility-first framework for satellite integration, designed to complement robustness-oriented core allocations such as those employed in SMDT-type architectures. The framework formalizes four nested feasibility layers:
\begin{itemize}
	\item \textit{Physical feasibility}, ensuring that satellite trades are admissible under concave market-impact laws;
	\item \textit{Economic feasibility}, suppressing noise-dominated reallocations via cost-dominance threshold constraints;
	\item \textit{Structural feasibility}, bounding satellite size through a structural optionality budget defined by tolerable loss if the thematic thesis fails;
	\item \textit{Epistemic feasibility}, limiting satellite breadth and dispersion via an entropy-based complexity budget.
\end{itemize}

Within this hierarchy, we argue that structural optionality provides the most natural primary design principle for thematic satellites. Satellites are treated not as continuously optimized return engines, but as bounded, slow-moving exposures representing long-horizon optionality on structural change. The remaining feasibility layers act as robustness lenses, ensuring that physical implementability, economic significance, and governance discipline are preserved.

Importantly, the framework is non-predictive by construction. It does not assume that thematic assets earn abnormal returns, nor does it rely on factor premia, timing signals, or backtested performance. Instead, it specifies the conditions under which a satellite allocation is admissible—in the sense that it does not undermine the robustness, interpretability, or survivability of the portfolio—irrespective of realized outcomes.

By separating feasibility from forecasting, this paper contributes a missing intermediate layer between high-level portfolio architecture and asset-level allocation, and provides a principled basis for integrating thematic satellites into small, robustness-oriented portfolios.

\subsection{Research gap and contribution}
\label{sec:gap}

Despite the practical ubiquity of core--satellite portfolio architectures, the academic
treatment of satellite integration remains weakly formalized, particularly for
\emph{small} portfolios and robustness-oriented allocation schemes. Existing approaches
typically assess satellite feasibility through one of three lenses: factor exposure and
optimization \citep{michaud1989markowitz,deMiguel2009}, thematic classification and
performance attribution \citep{bender2018thematic}, or liquidity and market-impact
capacity \citep{almgren2001optimal,toth2011}. These approaches implicitly assume that
feasibility is either ensured by construction or becomes binding only at institutional
scale. As a result, the literature leaves a conceptual gap between high-level portfolio
architecture and asset-level allocation: it lacks a \emph{theme-agnostic, scale-aware}
definition of satellite \emph{admissibility} that is independent of return forecasting.

This gap is especially salient in the small-portfolio regime that motivates the present
study. At modest scale, concave market-impact constraints are often non-binding for liquid
assets \citep{bouchaud2018trades}, while the dominant constraints arise elsewhere. Fixed
trading frictions and discrete execution imply that frequent micro-rebalancing is
economically dominated by noise rather than information
\citep{constantinides1986portfolio}. Thematic exposure introduces irreversibility risk,
making the cost of being wrong asymmetric and balance-sheet relevant
\citep{dixit1994investment}. In addition, portfolio breadth and dispersion are constrained
by governance, monitoring, and interpretability capacity rather than by diversification
theory alone \citep{gennaioli2018crash,cochrane2011}. Although these constraints are widely
recognized in practice, they are rarely integrated into a single, formal framework that
yields explicit, ex ante design bounds for satellite allocations.

This paper addresses that gap by reframing satellite integration as a
\emph{feasibility-first} problem rather than a predictive or optimization problem. We make
three distinct contributions. First, we reinterpret satellite allocations as
\emph{structural optionality} instead of scaled-down active strategies or thematic bets,
explicitly bounding exposure by tolerable loss under thesis failure, consistent with
theories of commitment and irreversibility under uncertainty
\citep{dixit1994investment}. Second, we introduce a hierarchical feasibility framework that
separates physical, economic, structural, and epistemic constraints, clarifying which
constraints bind at small portfolio scales and why, and reflecting robustness-oriented
critiques of optimization under limited action resolution and governance capacity
\citep{michaud1989markowitz,chorin2009}. Third, we derive closed-form feasibility bounds on
satellite size, turnover, and breadth that do not rely on alpha estimates, factor premia,
or backtested performance.

By separating feasibility from forecasting, the proposed framework provides a principled
intermediate layer between portfolio architecture and asset-level allocation. The approach
complements robustness-oriented core constructions, such as SMDT-style architectures, and
is compatible with domain conditioning based on institutional and informational continuity,
as captured by GAER-type diagnostics \citep{lo2004adaptive,shiller2003}. In this sense, the
contribution is not a new return model, but a clarification of when thematic satellites can
be integrated without undermining robustness, interpretability, or survivability.

\section{Related Work}

\subsection{Factor investing and robustness-oriented allocation}

A large literature studies factor-based portfolio construction, documenting cross-sectional return patterns associated with characteristics such as value, size, momentum, profitability, and investment. While factor models provide a powerful descriptive and organizational framework, their practical implementation has increasingly emphasized robustness, estimation error, and turnover control rather than pure mean--variance optimality.

This shift has motivated approaches such as equal-weighting, rank-based allocations, bounded tilts, and constraint-aware optimization, all of which seek to reduce sensitivity to parameter uncertainty and non-stationarity. Importantly, these methods implicitly assume a minimum degree of informational continuity and economic feasibility across the investable universe. However, the literature typically treats the domain of applicability of such methods as given, relying on liquidity screens or coverage thresholds rather than explicit feasibility diagnostics.

The present paper does not propose an alternative factor model, nor does it evaluate factor premia. Instead, it addresses a logically prior question: under what conditions can satellite allocations be added to a robustness-oriented core without violating feasibility constraints that factor models leave implicit?

\subsection{Thematic investing and thematic ETFs}

Thematic investing has grown rapidly, particularly through the proliferation of thematic ETFs targeting technologies, industries, or narratives such as artificial intelligence, robotics, clean energy, or genomics. The academic and practitioner literature on thematic strategies largely focuses on classification schemes, exposure mapping, and performance evaluation, often emphasizing long-horizon growth narratives or innovation cycles.

A recurring concern in this literature is that thematic portfolios tend to exhibit high concentration, elevated turnover, and sensitivity to capital flows. Empirical studies document that thematic ETF performance is often strongly correlated with investor flows, raising questions about whether observed returns reflect structural growth or transient demand effects. Moreover, thematic indices typically rebalance mechanically based on classification changes, introducing implicit trading rules that may be poorly aligned with portfolio scale or investor objectives.

Our framework differs fundamentally in perspective. We do not attempt to design or evaluate thematic portfolios as standalone products. Instead, we treat thematic exposure as a satellite component embedded within a broader portfolio architecture, whose primary role is to express bounded optionality rather than to maximize exposure to a narrative. From this standpoint, the relevant question is not whether a theme outperforms, but whether its inclusion is feasible given portfolio scale, governance capacity, and tolerance for irreversibility.

\subsection{Market impact, liquidity, and implementation constraints}

Market impact models and liquidity-based capacity constraints play a central role in the literature on portfolio implementation, particularly for large institutional investors. Concave impact laws—most prominently the square-root law—provide a well-established framework for bounding feasible trade sizes as a function of average daily volume and turnover.

While these results are essential at scale, their relevance diminishes sharply for small portfolios. At modest AUM levels, impact-based constraints are typically slack for liquid assets, and implementation feasibility is dominated by fixed costs, spreads, and governance considerations rather than by price impact. Nonetheless, impact models remain a necessary outer feasibility layer, ensuring physical admissibility of trades.

This paper incorporates market-impact constraints explicitly, but treats them as necessary but generally non-binding conditions in the small-portfolio regime. The contribution lies in identifying and formalizing the additional feasibility layers that become binding precisely when impact does not.

\subsection{Portfolio governance, complexity, and bounded rationality}

A smaller but growing literature emphasizes the role of governance, monitoring costs, and bounded rationality in portfolio management. From this perspective, overly complex portfolios may underperform not because of market frictions, but because they exceed the decision-maker’s capacity to monitor, interpret, and act consistently.

Entropy-based measures of portfolio concentration and diversification have been proposed as descriptive tools, but are rarely used as explicit constraints. We extend this idea by introducing an entropy budget as a feasibility condition, bounding the incremental complexity introduced by satellite allocations. This provides a formal link between portfolio structure and governance capacity, which is particularly relevant for small investors and decentralized decision environments.


\section{Generalized Satellite Design under Feasibility Discipline}
\label{sec:satellite_general}

\subsection{Objective and scope}
\label{sec:satellite_general_objective}

The objective of the satellite component is to capture \emph{structural adoption rents}
associated with a well-defined long-horizon theme (e.g., artificial intelligence,
energy transition, digital infrastructure, defense modernization, healthcare platforms),
while avoiding exposure to short-term hype, early-stage venture dynamics, or fragile
thematic beta.

The satellite is therefore designed as \emph{bounded structural optionality} rather than
as an actively optimized or signal-driven strategy. The emphasis is on feasibility,
institutional embedding, and persistence of cash-flow extraction, not on thematic
completeness or narrative purity.

\subsection{Allocation size and risk framing}
\label{sec:satellite_general_size}

The satellite allocation is bounded ex ante as a fraction of total portfolio value:
\begin{equation}
	\alpha \in [\alpha_{\min}, \alpha_{\max}],
	\qquad \alpha_{\max} \in [0.10,0.15],
\end{equation}
with the effective value of $\alpha$ determined by the structural optionality budget
(Section~\ref{sec:struct_feas}). Satellite sizing is \emph{risk-budgeted rather than
	conviction-weighted}: exposure reflects tolerable loss under thesis failure, not subjective
confidence in the theme.

\subsection{Eligible asset universe}
\label{sec:satellite_general_universe}

Let $\mathcal{T}$ denote a thematic hypothesis (e.g., increased AI compute intensity,
electrification, digitization of public services). The eligible satellite universe is
restricted to assets that satisfy both:
\begin{enumerate}
	\item \textbf{Thematic relevance}: the theme contributes materially to long-run demand,
	pricing power, or cost advantage;
	\item \textbf{Domain feasibility}: assets belong to the GAER-admissible domain,
	$\mathcal{U}_{\text{GAER}}$.
\end{enumerate}

Formally,
\begin{equation}
	\mathcal{S}_{\mathcal{T}} \subseteq \mathcal{U}_{\text{GAER}},
\end{equation}
ensuring that the satellite does not attempt to compensate for adverse institutional,
regulatory, or geopolitical conditions through thematic exposure.

\subsection{Generalized tier structure}
\label{sec:satellite_general_tiers}

To preserve feasibility purity and governance clarity, the satellite is organized into
tiers reflecting \emph{where and how} structural rents are extracted along the thematic
value chain. The tiering principle is generic and applies to any theme.

\paragraph{Tier A — Hard constraints and bottlenecks.}
Assets controlling capital-intensive, scarce, or regulated bottlenecks that are necessary
for the theme to scale. These entities typically exhibit high barriers to entry, systemic
relevance, and strong institutional embedding.

\paragraph{Tier B — Platforms and rent extractors.}
Assets that intermediate, orchestrate, or monetize the theme through platforms, networks,
or long-duration contractual relationships. Rents arise from pricing power, lock-in, or
ecosystem control rather than from speculative end-demand.

\paragraph{Tier C — Embedded adopters.}
Assets for which the theme enhances margins, efficiency, or switching costs without being
the primary revenue driver. Adoption is gradual, persistent, and complementary to existing
business models.

This tiered decomposition abstracts from sector labels and instead reflects the economic
role played by each asset within the thematic diffusion process.

\subsection{Internal weighting rules}
\label{sec:satellite_general_weights}

Internal satellite weights are determined by parsimonious, rule-based schemes rather than
optimization. Let $K_A$, $K_B$, and $K_C$ denote the number of constituents in Tiers A, B,
and C, respectively, with $K = K_A + K_B + K_C$.

Weights are assigned as:
\begin{equation}
	w_i =
	\begin{cases}
		\frac{\alpha}{K} \cdot \kappa_A, & i \in \text{Tier A},\\[6pt]
		\frac{\alpha}{K}, & i \in \text{Tier B},\\[6pt]
		\frac{\alpha}{K} \cdot \kappa_C, & i \in \text{Tier C},
	\end{cases}
\end{equation}
where $\kappa_A \ge 1$, $\kappa_C \le 1$, and normalization ensures $\sum_i w_i = \alpha$.
This implements equal-weighting within tiers with an optional mild tilt toward upstream
constraints and away from downstream adopters.

\subsection{Rebalancing discipline}
\label{sec:satellite_general_rebalancing}

Rebalancing is intentionally infrequent and governance-driven. The default schedule is
annual or semi-annual. Rebalancing is \emph{not} triggered by price momentum, valuation
signals, or short-term thematic news.

Adjustments are permitted only upon identifiable \emph{structural breaks}, including:
\begin{itemize}
	\item regulatory or policy regime shifts affecting the theme;
	\item supply-side discontinuities or constraint realignments;
	\item institutional or geopolitical changes altering GAER admissibility;
	\item technological standardization events that relocate rents across tiers.
\end{itemize}

Absent such events, the satellite remains unchanged.

\subsection{Explicit exclusions}
\label{sec:satellite_general_exclusions}

Across themes, certain asset categories are excluded from satellite eligibility by design,
independently of expected returns or thematic relevance.
Table~\ref{tab:satellite_exclusions} summarizes the asset categories that are excluded from
satellite eligibility based on feasibility considerations.

\begin{table}[H]
	\centering
	\small
	\setlength{\tabcolsep}{6pt}
	\renewcommand{\arraystretch}{1.15}
	\begin{tabular}{ll}
		\toprule
		\textbf{Excluded category} & \textbf{Rationale} \\
		\midrule
		Pure-play early-stage firms &
		Venture-like risk and regime fragility \\
		
		Small-cap thematic specialists &
		Liquidity, disclosure, and impact constraints \\
		
		Regime-opaque jurisdictions &
		Institutional and geopolitical discontinuity \\
		
		Thematic ETFs &
		Factor dilution and mechanically induced turnover \\
		\bottomrule
	\end{tabular}
	
	\caption{Asset categories excluded from thematic satellite eligibility.
		Exclusions are based on feasibility considerations rather than return
		expectations, reflecting constraints related to liquidity, governance,
		and regime stability.}
	\label{tab:satellite_exclusions}
\end{table}

These exclusions reinforce the principle that the satellite is not a vehicle for thematic
completeness, but a controlled expression of feasibility-consistent optionality.

\subsection{Interpretation}

The generalized satellite framework applies uniformly across themes by focusing on
\emph{economic role} rather than narrative labels. By combining GAER-based domain
conditioning with tiered exposure, bounded optionality, and disciplined rebalancing,
the satellite remains compatible with robustness-oriented core allocations and preserves
the interpretability and survivability of the overall portfolio.


\section{Worked Examples: Applying the Tiered Satellite Framework}
\label{sec:worked_examples}

This section provides two worked examples illustrating how the generalized, feasibility-first
satellite framework applies to distinct structural themes. The examples are intentionally
non-exhaustive and non-optimizing, and are presented to demonstrate transferability of the
tier logic across domains.

\subsection{Artificial Intelligence}
\label{sec:worked_example_ai}

\subsubsection{Thematic hypothesis}

The artificial intelligence theme is defined narrowly as the long-horizon diffusion of
compute-intensive and data-driven technologies across production, services, and public
infrastructure. The objective is to capture \emph{structural adoption rents} arising from
bottlenecks, platform control, and switching costs, rather than short-term valuation
re-rating or speculative innovation cycles.

\subsubsection{Eligible universe and GAER conditioning}

The AI satellite universe is restricted to AI \emph{enablers} that satisfy GAER domain
admissibility:
\begin{equation}
	\mathcal{S}_{\text{AI}} \subseteq \mathcal{U}_{\text{GAER}}.
\end{equation}
This excludes firms whose AI exposure depends on fragile funding conditions, opaque
disclosure, or discontinuous regulatory regimes.

\subsubsection{Tiered decomposition}

\paragraph{Tier A — Compute and infrastructure bottlenecks.}
This tier includes firms controlling capital-intensive and scarce components required for
AI scaling, such as advanced semiconductors, lithography, and foundry capacity. Rents arise
from monopoly-like positions, extreme capital intensity, and systemic relevance. These
assets are typically embedded within U.S., EU, or allied institutional cores and benefit
from implicit backstops in stress scenarios.

\paragraph{Tier B — Platforms and rent extractors.}
This tier includes firms monetizing AI through cloud platforms, enterprise software stacks,
and ecosystem control. Rents are extracted via pricing power over compute, APIs, and
long-duration enterprise contracts rather than end-user hype.

\paragraph{Tier C — Embedded enterprise adopters.}
This tier includes firms for which AI adoption increases switching costs, margins, or
operational efficiency without being the primary revenue driver. Adoption is gradual and
persistent, aligning well with low-turnover satellite design.

\subsubsection{Sizing, weights, and rebalancing}

Satellite size $\alpha_{\text{AI}}$ is determined by the structural optionality budget:
\begin{equation}
	\alpha_{\text{AI}} \le \frac{L}{D_{\max}},
\end{equation}
with internal weights assigned by equal-weighting within tiers and an optional mild tilt
favoring Tier~A over Tier~C. Rebalancing is annual or semi-annual and occurs only upon
identifiable structural breaks (e.g., export controls, foundry access, or regulatory regime
changes). Price momentum and short-term AI news do not trigger rebalancing.

\subsubsection{Explicit exclusions}

Excluded from the AI satellite are pure-play AI startups, small-cap AI software firms,
AI-themed ETFs, and AI exposures in geopolitically discontinuous jurisdictions. These
exclusions reflect feasibility purity rather than thematic completeness.

\subsection{Defense and Security Modernization}
\label{sec:worked_example_defense}

\subsubsection{Thematic hypothesis}

The defense theme is defined as the long-horizon modernization of military, security, and
dual-use infrastructure driven by geopolitical realignment, alliance commitments, and
technological upgrading. The objective is to capture persistent rents associated with
defense procurement, systems integration, and embedded security requirements, rather than
cyclical conflict escalation or short-term budget announcements.

\subsubsection{Eligible universe and GAER conditioning}

Defense-related candidates are restricted to GAER-admissible jurisdictions characterized by
stable procurement processes, transparent budgeting, and alliance-based security
commitments:
\begin{equation}
	\mathcal{S}_{\text{DEF}} \subseteq \mathcal{U}_{\text{GAER}}.
\end{equation}
This excludes defense exposure in opaque regimes, ad hoc national champions, or jurisdictions
with discontinuous contract enforcement.

\subsubsection{Tiered decomposition}

\paragraph{Tier A — Hard constraints and strategic bottlenecks.}
This tier includes firms controlling capital-intensive and strategically critical assets,
such as prime contractors, shipyards, aerospace manufacturing capacity, and munitions
production. Rents arise from long-duration contracts, limited competition, and systemic
importance to national security.

\paragraph{Tier B — Systems integrators and platforms.}
This tier includes firms specializing in systems integration, avionics, command-and-control,
cybersecurity, and ISR (intelligence, surveillance, reconnaissance) platforms. Rents are
derived from integration complexity, certification barriers, and long-term service
agreements.

\paragraph{Tier C — Embedded defense and security suppliers.}
This tier includes firms providing software, logistics, maintenance, and dual-use
technologies embedded within broader defense systems. Adoption is incremental and persistent,
supporting low-turnover satellite structures.

\subsubsection{Sizing, weights, and rebalancing}

Satellite size $\alpha_{\text{DEF}}$ is set via the same structural optionality budget used
for other themes:
\begin{equation}
	\alpha_{\text{DEF}} \le \frac{L}{D_{\max}}.
\end{equation}
Internal weights follow equal-weighting within tiers with an optional tilt toward Tier~A.
Rebalancing is infrequent and triggered only by structural changes, such as alliance
reconfiguration, procurement regime shifts, or institutional realignments affecting GAER
admissibility.

\subsubsection{Explicit exclusions}

Excluded from the defense satellite are small-cap defense specialists with limited contract
visibility, speculative cybersecurity startups, firms reliant on single-country conflict
exposure, and defense-themed ETFs. These exclusions avoid liquidity risk, governance opacity,
and mechanically induced turnover.

\subsection{Cross-theme interpretation}

Across both AI and defense, the same feasibility-first logic applies: exposure is organized
around constraint location, institutional embedding, and rent persistence rather than
sector labels or narratives. This demonstrates that the tiered satellite framework is
theme-agnostic and can be applied uniformly to structurally distinct domains while remaining
compatible with SMDT-style robustness and GAER-based domain conditioning.


\section{Four Feasibility Approaches for Satellite Integration}
\label{sec:feasibility-four}

\begin{figure}[H]
	\centering
	\begin{tikzpicture}[
		scale=0.82,
		transform shape,
		node distance=8mm,
		every node/.style={
			draw,
			rectangle,
			rounded corners,
			align=center,
			minimum width=9cm,
			minimum height=8mm
		},
		arrow/.style={->, thick}
		]
		
		\node (physical) {%
			\textbf{Physical feasibility}\\
			\small Impact admissibility under concave market--impact laws
		};
		
		\node (economic) [above=of physical] {%
			\textbf{Economic feasibility (Cost-Dominance Threshold)}\\
			\small Suppression of noise-dominated and friction-driven reallocations
		};
		
		\node (structural) [above=of economic] {%
			\textbf{Structural feasibility (Optionality budget)}\\
			\small Bounding exposure by tolerable loss under thematic failure
		};
		
		\node (epistemic) [above=of structural] {%
			\textbf{Epistemic feasibility (Entropy / governance)}\\
			\small Limiting breadth and dispersion to preserve interpretability
		};
		
		\draw[arrow] (physical.north) -- (economic.south);
		\draw[arrow] (economic.north) -- (structural.south);
		\draw[arrow] (structural.north) -- (epistemic.south);
		
	\end{tikzpicture}
	
	\caption{Feasibility hierarchy for satellite integration.
		Satellite admissibility is evaluated through a sequence of nested constraints:
		physical feasibility, economic feasibility (Cost-Dominance Threshold),
		structural feasibility (optionality budget), and epistemic feasibility
		(entropy and governance). A satellite allocation is admissible only if all
		lower-level constraints are satisfied.}
	\label{fig:feasibility_hierarchy}
\end{figure}
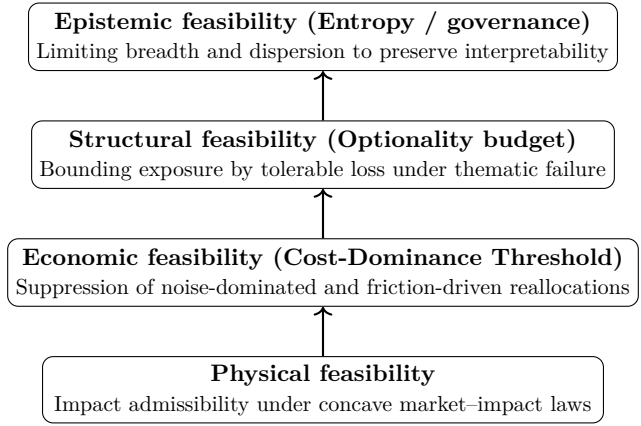

Satellite integration is formulated as a feasibility problem rather than a forecasting or factor-extraction problem. We introduce four complementary approaches that operate at distinct layers of admissibility. A satellite allocation is considered admissible only if it satisfies all four criteria; the binding criterion depends on portfolio scale and governance context. Figure~\ref{fig:feasibility_hierarchy} summarizes the hierarchical structure of feasibility constraints used to assess satellite admissibility.

\subsection{Approach I: Physical Feasibility (Market Impact)}
\label{sec:phys_feas}

\textbf{Purpose.} Ensure that satellite trades are physically implementable under concave market-impact laws. This layer is essential at scale, and provides an outer admissibility envelope at any scale.

\textbf{Primitive constraint.} Let $V_i$ denote dollar ADV and $Q_i$ the dollar notional traded in asset $i$ at rebalancing. Assume a concave impact law
\begin{equation}
	I_i \;=\; c\left(\frac{Q_i}{V_i}\right)^{\delta}, \qquad 0<\delta<1,
\end{equation}
with $Q_i \approx A\,w_i\,\tau_i$, where $A$ is AUM, $w_i$ is the target weight, and $\tau_i\in[0,1]$ is turnover fraction at the rebalance.

\textbf{Feasibility bound.} Imposing $I_i \le \bar I_i$ yields
\begin{equation}
	w_i \;\le\; \frac{V_i}{A\,\tau_i}\left(\frac{\bar I_i}{c}\right)^{1/\delta}.
\end{equation}
A commonly used specialization is a participation cap $Q_i/V_i \le \phi$, implying
\begin{equation}
	w_i \;\le\; \frac{\phi\,V_i}{A\,\tau_i}.
\end{equation}

\subsection{Approach II: Economic Feasibility (Cost-Dominance Threshold)}
\label{sec:econ_feas}

\textbf{Purpose.} Suppress rebalancing actions whose economic effect is dominated by fixed frictions and noise, thereby controlling turnover and ticket fragmentation in small portfolios.

\textbf{Primitive constraint.} Let $C_{\mathrm{rt}}$ denote round-trip trading costs (in bps) for the relevant trading path (spread, commissions, and any FX conversions), and let $\varepsilon$ denote a minimum economically meaningful portfolio effect (in bps). A weight change $\Delta w$ is admissible only if
\begin{equation}
	\Delta w\,C_{\mathrm{rt}} \;\ge\; \varepsilon.
\end{equation}

\textbf{Feasibility bounds.} Define the minimum meaningful weight change
\begin{equation}
	\Delta w_{\min} \;=\; \frac{\varepsilon}{C_{\mathrm{rt}}}.
\end{equation}
For a satellite sleeve of size $\alpha$ (fraction of total portfolio), this implies a natural upper bound on satellite breadth (coarse-graining):
\begin{equation}
	K \;\le\; \frac{\alpha}{\Delta w_{\min}},
\end{equation}
under the design rule that each active satellite constituent must carry at least $\Delta w_{\min}$ weight (within the total portfolio) to avoid noise-dominated trades.

\subsection{Approach III: Structural Feasibility (Optionality Budget)}
\label{sec:struct_feas}

\textbf{Purpose.} Bound satellite size by balance-sheet tolerance under thesis failure. This approach treats the satellite as bounded, slow-moving structural optionality rather than a continuously optimized return engine.

\textbf{Primitive constraint.} Let $\alpha$ denote the satellite share of the total portfolio, $D_{\max}$ a conservative maximum plausible drawdown of the satellite under thesis failure, and $L$ the maximum acceptable total-portfolio loss attributable to the satellite.

\textbf{Feasibility bound.} The structural optionality budget is
\begin{equation}
	\alpha\,D_{\max} \;\le\; L
	\qquad \Longrightarrow \qquad
	\alpha \;\le\; \frac{L}{D_{\max}}.
\end{equation}
Once $\alpha$ is fixed, internal satellite weights can be chosen by parsimonious rules (e.g., equal-weight across a small set) to preserve interpretability and reduce model dependence.

\subsection{Approach IV: Epistemic Feasibility (Entropy / Complexity Budget)}
\label{sec:epi_feas}

\textbf{Purpose.} Limit incremental complexity (monitoring, interpretability, and governance burden) introduced by satellites. This is particularly relevant for small portfolios where decision capacity is a binding resource.

\textbf{Primitive constraint.} Define portfolio weight entropy
\begin{equation}
	H \;=\; -\sum_{j=1}^{N} w_j \log w_j,
\end{equation}
and impose an entropy increment budget $\Delta H_{\max}$:
\begin{equation}
	H_{\mathrm{total}} - H_{\mathrm{core}} \;\le\; \Delta H_{\max}.
\end{equation}

\textbf{Feasibility bound.} For a satellite of total weight $\alpha$ distributed equally across $K$ names, the incremental entropy is approximated by
\begin{equation}
	\Delta H_{\mathrm{sat}} \approx -\alpha \log\!\left(\frac{\alpha}{K}\right).
\end{equation}
Imposing $\Delta H_{\mathrm{sat}} \le \Delta H_{\max}$ yields an upper bound on breadth:
\begin{equation}
	K \;\le\; \alpha \exp\!\left(\frac{\Delta H_{\max}}{\alpha}\right).
\end{equation}

\subsection{Comparative Summary}
\label{sec:feasibility-compare}

Table~\ref{tab:feasibility-compare} compares the four approaches by constraint
type, required inputs, binding regime, and typical design implication for
small core--satellite portfolios.

\begin{table}[H]
	\centering
	\footnotesize
	\setlength{\tabcolsep}{5pt}
	\renewcommand{\arraystretch}{1.15}
	\begin{tabularx}{\linewidth}{@{}
			p{0.15\linewidth}
			p{0.18\linewidth}
			p{0.20\linewidth}
			p{0.20\linewidth}
			Y
			@{}}
		\toprule
		\textbf{Approach} &
		\textbf{Feasibility layer} &
		\textbf{Primitive inputs} &
		\textbf{Typically binding when} &
		\textbf{Main design implication} \\
		\midrule
		I. Market impact &
		Physical implementability &
		ADV $V_i$; turnover $\tau_i$; AUM $A$; impact cap $\bar I$ (or participation $\phi$) &
		Large AUM and/or high turnover; thin liquidity &
		Liquidity-weighted caps on $w_i$; penalize frequent rebalancing \\
		\addlinespace
		
		II. Cost-Dominance Threshold &
		Economic significance of trades &
		Round-trip costs $C_{\mathrm{rt}}$; minimum effect $\varepsilon$; sleeve size $\alpha$ &
		Small portfolios; fragmented tickets; high relative frictions &
		Coarse-grained satellites; upper bound on $K$; suppress micro-rebalances \\
		\addlinespace
		
		III. Optionality budget &
		Balance-sheet survivability &
		Satellite share $\alpha$; thesis-failure drawdown $D_{\max}$; loss tolerance $L$ &
		Whenever thesis risk is salient; preference for explicit loss bounds &
		Explicit upper bound on $\alpha$; satellites interpreted as bounded optionality \\
		\addlinespace
		
		IV. Entropy budget &
		Epistemic / governance capacity &
		Entropy increment $\Delta H_{\max}$; sleeve size $\alpha$; breadth $K$ &
		Limited monitoring capacity; desire for interpretability &
		Upper bound on $K$ and dispersion; discourages pseudo-diversification \\
		\bottomrule
	\end{tabularx}
	
	\caption{Comparison of four feasibility approaches for satellite integration.
		The approaches are complementary and may be treated as nested admissibility layers:
		physical $\rightarrow$ economic $\rightarrow$ structural $\rightarrow$ epistemic.}
	\label{tab:feasibility-compare}
\end{table}

Each feasibility layer imposes a distinct design constraint and excludes
specific classes of assumptions that are common in predictive or
optimization-based approaches.
Table~\ref{tab:feasibility_mapping} provides a concise mapping between
feasibility layers, their economic interpretation, and the portfolio
dimensions they constrain.

\begin{table}[H]
	\centering
	\footnotesize
	\setlength{\tabcolsep}{5pt}
	\renewcommand{\arraystretch}{1.15}
	\begin{tabular}{
			p{2.8cm}
			p{3.6cm}
			p{3.2cm}
			p{3.8cm}
		}
		\toprule
		\textbf{Feasibility layer} &
		\textbf{Economic interpretation} &
		\textbf{What it bounds} &
		\textbf{What it does not assume} \\
		\midrule
		Physical feasibility &
		Impact admissibility under concave market impact &
		Trade size per rebalance &
		Return predictability or alpha \\
		
		Economic feasibility &
		Noise suppression via cost dominance &
		Turnover and action frequency &
		Timing skill or signal strength \\
		
		Structural feasibility &
		Irreversibility and balance-sheet risk &
		Maximum satellite weight &
		Growth forecasts or theme success \\
		
		Epistemic feasibility &
		Governance and interpretability capacity &
		Number of names and dispersion &
		Optimal diversification or factor structure \\
		\bottomrule
	\end{tabular}
	
	\caption{Interpretation of feasibility layers and associated design bounds.
		The table summarizes the economic meaning of each feasibility layer, the
		portfolio dimension it constrains, and the assumptions it explicitly does not
		rely on.}
	\label{tab:feasibility_mapping}
\end{table}

\subsection{Alternative criteria and minimal completeness}
\label{sec:feasibility_completeness}

A natural question is whether alternative feasibility criteria could replace or dominate
the hierarchical framework proposed in this paper. The portfolio construction literature
offers several candidate approaches, including liquidity- or capacity-based sizing rules
\citep{almgren2001optimal,toth2011}, volatility or drawdown-based risk budgets
\citep{deMiguel2009,clarke2013risk}, belief-weighted or Bayesian exposure schemes
\citep{pastor2015learning}, and governance heuristics based on discretionary limits.
Each of these addresses a subset of the constraints relevant to satellite integration.

Liquidity- and impact-based criteria provide necessary conditions for physical
implementability, but are typically non-binding at the small portfolio scales considered
here and do not address economic significance, irreversibility risk, or governance
capacity \citep{bouchaud2018trades}. Volatility- or tail-risk budgets quantify stochastic
variability but conflate reversible market fluctuations with structural loss under thematic
failure \citep{dixit1994investment}. Belief- or scenario-weighted approaches reintroduce
forecasting assumptions and subjective priors, undermining robustness and auditability
\citep{michaud1989markowitz,lo2004adaptive}. Informal governance heuristics capture
practical constraints but lack formal structure and cannot be translated into explicit,
auditable design bounds \citep{gennaioli2018crash}.

The feasibility hierarchy developed in this paper does not reject these perspectives;
rather, it subsumes them where appropriate. Physical feasibility encompasses standard
liquidity and market-impact constraints \citep{almgren2001optimal,toth2011}. Economic
feasibility formalizes the suppression of noise-dominated actions implied by fixed trading
frictions \citep{constantinides1986portfolio}. Structural feasibility captures
irreversibility and balance-sheet risk in a model-free manner consistent with real-options
logic \citep{dixit1994investment}. Epistemic feasibility renders governance and
interpretability constraints explicit through a complexity budget, in line with concerns
about belief fragility and overextension \citep{gennaioli2018crash,cochrane2011}. Within
the stated scope, alternative criteria either collapse to one of these layers or introduce
additional assumptions—most commonly predictive beliefs—that the present framework
deliberately excludes.

This observation motivates the following proposition.

\begin{proposition}[Minimal completeness of the feasibility hierarchy]
	\label{prop:minimal_completeness}
	Consider the class of satellite-integration frameworks that satisfy the following
	conditions: (i) feasibility is assessed ex ante and independently of return forecasting
	\citep{michaud1989markowitz}; (ii) admissibility is scale-aware and applicable in the
	small-portfolio regime \citep{bouchaud2018trades}; (iii) structural loss under thematic
	failure is bounded explicitly \citep{dixit1994investment}; and (iv) governance and
	interpretability constraints are treated as first-order design considerations
	\citep{gennaioli2018crash}.  
	
	Within this class, any feasibility criterion for satellite allocations can be represented
	as, or reduced to, a combination of physical, economic, structural, and epistemic
	constraints as defined in this paper. No strictly weaker subset of these constraints yields
	the same admissibility guarantees, and any strictly stronger framework necessarily
	introduces additional assumptions beyond feasibility, such as predictive beliefs or
	model-dependent valuation \citep{lo2004adaptive,shiller2003}.
\end{proposition}

\noindent
Proposition~\ref{prop:minimal_completeness} does not claim optimality in a return-maximizing
sense. Rather, it characterizes the feasibility hierarchy as a minimally complete design
layer: removing any constraint weakens admissibility guarantees, while adding further
structure moves the framework beyond feasibility and into forecasting or optimization.


\section{Integrated Feasibility Strategy and Compatibility with SMDT--GAER}
\label{sec:integrated_strategy}

This section operationalizes satellite integration as a multi-layer feasibility problem.
Rather than selecting satellites via return forecasts, factor premia, or backtested
performance, we treat satellite exposure as \emph{bounded optionality} subject to four
nested admissibility constraints: physical, economic, structural, and epistemic. The
resulting strategy is compatible with the SMDT architecture and with GAER-based universe
conditioning, because it preserves the separation between (i) informational feasibility
(domain), (ii) portfolio feasibility (implementation), and (iii) allocation rules (weights).

\subsection{Strategy overview: a four-layer admissibility cascade}
\label{sec:admissibility_cascade}

Let the total portfolio AUM be $A$ and let the satellite sleeve weight be $\alpha \in (0,1)$,
so that satellite AUM is $A_s = \alpha A$. Let $\mathcal{S}$ denote the candidate satellite
set, with weights $\{w_i\}_{i\in\mathcal{S}}$ summing to $\alpha$.

We implement the satellite through a cascade of feasibility filters:
\begin{enumerate}
	\item \textbf{GAER domain constraint (informational admissibility).}
	Restrict candidates to a GAER-admissible domain $\mathcal{U}_{\text{GAER}}$ (e.g., stable
	institutional embedding, disclosure/enforcement, geopolitical continuity). The satellite
	is defined \emph{within} this domain rather than used to redefine the domain.
	
	\item \textbf{Structural feasibility (primary sizing).}
	Choose $\alpha$ using an explicit structural optionality budget that bounds loss under
	thesis failure.
	
	\item \textbf{Epistemic feasibility (complexity budgeting).}
	Given $\alpha$, bound breadth $K = |\mathcal{S}|$ and dispersion by an entropy budget.
	
	\item \textbf{Economic feasibility (action resolution).}
	Enforce Cost-Dominance Threshold to suppress noise-dominated reallocations and
	fragmented tickets.
	
	\item \textbf{Physical feasibility (impact validation).}
	Verify that concave market-impact bounds are satisfied for all planned trades; at small
	scale this is typically slack but remains a necessary outer admissibility check.
\end{enumerate}

The ordering above is intentional: for small portfolios, structural and epistemic bounds
typically dominate; economic bounds discipline turnover; physical bounds provide a final
outer envelope.

\subsection{Layer definitions and binding constraints}
\label{sec:layer_defs}

\paragraph{(0) GAER conditioning (domain).}
Let $\mathcal{U}$ be the investable universe and $\mathcal{U}_{\text{GAER}}\subseteq\mathcal{U}$
the GAER-admissible subset. Satellite candidates must satisfy
\begin{equation}
	\mathcal{S} \subseteq \mathcal{U}_{\text{GAER}}.
\end{equation}
This preserves the interpretation of GAER as a \emph{domain feasibility diagnostic} rather
than a return signal.

\paragraph{(1) Structural optionality budget (sizing).}
Let $D_{\max}$ be a conservative maximum drawdown of the satellite sleeve under thesis
failure, and let $L$ be the maximum acceptable loss contribution (fraction of total
portfolio) attributable to the satellite. Require
\begin{equation}
	\alpha D_{\max} \le L
	\qquad \Longrightarrow \qquad
	\boxed{\alpha \le \frac{L}{D_{\max}}.}
\end{equation}
This yields a transparent upper bound on thematic exposure independent of liquidity,
turnover, or predictive assumptions.

\paragraph{(2) Entropy / complexity budget (breadth and dispersion).}
Define weight entropy $H = -\sum_j w_j \log w_j$ and impose an incremental budget
$\Delta H_{\max}$ for the satellite:
\begin{equation}
	H_{\text{total}} - H_{\text{core}} \le \Delta H_{\max}.
\end{equation}
For a satellite of total weight $\alpha$ spread equally across $K$ names,
\begin{equation}
	\Delta H_{\text{sat}} \approx -\alpha \log\!\left(\frac{\alpha}{K}\right).
\end{equation}
Imposing $\Delta H_{\text{sat}} \le \Delta H_{\max}$ yields
\begin{equation}
	\boxed{K \le \alpha \exp\!\left(\frac{\Delta H_{\max}}{\alpha}\right).}
\end{equation}
This constraint formalizes governance capacity and discourages pseudo-diversification in
small portfolios.

\paragraph{(3) Cost-Dominance Threshold (turnover control).}
Let $C_{\mathrm{rt}}$ denote round-trip trading costs (bps) and let $\varepsilon$ denote a
minimum economically meaningful portfolio effect (bps). A weight change $\Delta w$ is
admissible only if
\begin{equation}
	\Delta w\,C_{\mathrm{rt}} \ge \varepsilon
	\qquad \Longrightarrow \qquad
	\boxed{\Delta w_{\min} = \frac{\varepsilon}{C_{\mathrm{rt}}}.}
\end{equation}
This suppresses frequent micro-rebalances and forces coarse-grained, auditable actions.
Given sleeve size $\alpha$, a sufficient practical bound on breadth is
\begin{equation}
	\boxed{K \le \frac{\alpha}{\Delta w_{\min}}}
\end{equation}
under the design rule that each active satellite constituent carries at least
$\Delta w_{\min}$ total-portfolio weight.

\paragraph{(4) Market-impact admissibility (physical validation).}
Let $V_i$ denote ADV and $Q_i$ traded notional at rebalance. Under concave impact
\begin{equation}
	I_i = c\left(\frac{Q_i}{V_i}\right)^{\delta}, \qquad 0<\delta<1,
\end{equation}
with $Q_i \approx A\,|w_i^{\text{new}}-w_i^{\text{old}}|$ (or $A w_i \tau_i$ as an envelope),
imposing $I_i \le \bar I_i$ yields
\begin{equation}
	\boxed{
		w_i \le \frac{V_i}{A\,\tau_i}\left(\frac{\bar I_i}{c}\right)^{1/\delta},
	}
\end{equation}
or, under a participation cap $Q_i/V_i \le \phi$,
\begin{equation}
	\boxed{
		w_i \le \frac{\phi V_i}{A\,\tau_i}.
	}
\end{equation}

\subsection{Operational procedure (algorithmic template)}
\label{sec:operational_procedure}

The integrated strategy can be implemented with the following template:
\begin{enumerate}
	\item \textbf{Domain selection.} Construct $\mathcal{U}_{\text{GAER}}$ and restrict the
	candidate satellite list accordingly.
	
	\item \textbf{Set the optionality budget.} Choose $(L, D_{\max})$ and set
	$\alpha = \min\{\alpha_{\text{policy}}, L/D_{\max}\}$, where $\alpha_{\text{policy}}$ is an
	ex ante policy cap.
	
	\item \textbf{Set complexity budget.} Choose $\Delta H_{\max}$ and derive an admissible
	breadth $K$; specify a parsimonious weighting rule (e.g., equal-weight, capped weights).
	
	\item \textbf{Set action resolution.} Choose $(C_{\mathrm{rt}}, \varepsilon)$ and compute
	$\Delta w_{\min}$; enforce no-trade regions and minimum ticket sizes consistent with
	$\Delta w_{\min}$.
	
	\item \textbf{Validate impact.} Using ADV estimates $V_i$, verify that planned trades satisfy
	impact/participation bounds. If any name fails, adjust $K$, reduce $\alpha$, or replace
	the name with a higher-ADV proxy.
\end{enumerate}

\subsection{Compatibility with SMDT}
\label{sec:compat_smdt}

The four-layer strategy is compatible with SMDT because it respects SMDT's separation
between \emph{robust allocation rules} and \emph{feasibility constraints}. SMDT-type cores
typically rely on low-estimation-error constructions (e.g., equal-weight baselines and
bounded tilts) and treat turnover as a central friction. The present satellite framework
does not introduce predictive signals into the core; it adds an outer governance layer
that:
(i) bounds satellite size via explicit loss tolerance (structural optionality),
(ii) bounds complexity (entropy),
(iii) bounds turnover via Cost-Dominance Threshold, and
(iv) verifies physical implementability via impact laws.
Therefore, the satellite remains an add-on that does not alter the internal logic of the
core weighting scheme.

\subsection{Compatibility with GAER}
\label{sec:compat_gaer}

GAER is interpreted as a \emph{domain conditioning diagnostic} that characterizes the
institutional and informational environment in which robust, rank-based, or bounded-tilt
strategies remain feasible. The present satellite framework is compatible with GAER
because it:
\begin{enumerate}
	\item Treats GAER as an \emph{ex ante admissibility filter} defining $\mathcal{U}_{\text{GAER}}$,
	not as an expected-return signal or timing device;
	\item Applies satellite feasibility constraints \emph{within} $\mathcal{U}_{\text{GAER}}$,
	thereby avoiding the misinterpretation of the satellite as a mechanism to ``override''
	domain feasibility;
	\item Reinforces the core implication of GAER: the persistence of informational signals
	and the continuity of price discovery are prerequisites for robustness-oriented
	allocations.
\end{enumerate}

\subsection{Discussion: which constraints bind in small portfolios?}
\label{sec:binding_discussion}

In small core--satellite portfolios, market-impact bounds derived from concave impact laws
are typically slack for liquid assets, and thus serve primarily as a validation layer.
By contrast, (i) cost-dominance threshold constraints, (ii) structural optionality
budgets, and (iii) entropy budgets are often binding, because they encode the practical
dominants at small scale: fixed frictions, tolerable loss under thesis failure, and
governance capacity. This explains why feasibility-first satellite integration can remain
non-predictive while still producing actionable and auditable design bounds on satellite
size, turnover, and breadth.

\subsection{Scope and limitations}
\label{sec:scope_limitations}

The framework developed in this paper is intentionally non-predictive and does not make
claims about the return performance of thematic satellites relative to benchmarks or
alternative allocation schemes. Parameter choices such as impact tolerances, loss budgets,
entropy limits, or cost-dominance threshold parameters are treated as policy inputs rather
than estimated quantities, and are therefore not calibrated within the paper. The worked
examples are illustrative mappings of the proposed feasibility logic to representative
themes and should not be interpreted as security recommendations or as evidence of
outperformance. Finally, while the analysis is motivated by small core--satellite
portfolios, the framework does not seek to characterize optimal behavior at large
institutional scale, where different constraints may bind. These limitations are
deliberate: the contribution of the paper is to formalize a missing feasibility layer for
satellite integration, not to propose a new predictive model or optimization criterion.

\appendix
\section{Calibration Notes and Design Parameters}
\label{app:calibration}

This appendix clarifies the role of calibration choices in the proposed feasibility-first
framework. All parameters introduced in the main text are treated as \emph{design and
	governance inputs}, not as estimated quantities, and are therefore not calibrated using
historical return data, factor regressions, or optimization criteria.

\subsection{Impact and participation parameters}

Parameters such as the impact tolerance $\bar I$, the participation cap $\phi$, and the
impact exponent $\delta$ enter the physical feasibility layer as \emph{outer admissibility
	checks}. At the portfolio scales motivating this study, these constraints are typically
slack for liquid assets and are included primarily to ensure physical implementability
rather than to optimize execution costs. Reasonable parameter ranges can be informed by
execution practice or market conventions, but precise calibration is not required for the
framework to operate.

\subsection{Cost-Dominance Threshold parameters}

The cost-dominance threshold parameters $\varepsilon$ and round-trip cost proxy
$C_{\mathrm{rt}}$ define the economic resolution of admissible trades. These parameters are
interpreted as governance tolerances reflecting the decision-maker’s willingness to act in
the presence of fixed frictions and noise. In practice, they can be set conservatively to
suppress frequent micro-adjustments; their role is to induce coarse-grained behavior rather
than to fine-tune performance.

\subsection{Structural optionality budgets}

The loss tolerance $L$ and maximum plausible drawdown $D_{\max}$ used to size satellite
exposure are policy choices reflecting balance-sheet risk preferences and irreversibility
concerns under thesis failure. They are not intended to approximate probabilistic risk
measures such as volatility or value-at-risk. Different investors may select different
values depending on mandate and horizon; the framework is agnostic to these choices and
simply translates them into explicit bounds on satellite size.

\subsection{Entropy and complexity budgets}

The entropy increment $\Delta H_{\max}$ is introduced as a proxy for governance,
interpretability, and monitoring capacity. It does not represent informational efficiency
in a statistical sense, nor does it aim to optimize diversification. Instead, it provides a
transparent mechanism to bound satellite breadth and dispersion relative to portfolio scale.
As with other parameters, $\Delta H_{\max}$ is selected ex ante and held fixed over time.

\subsection{Interpretation}

Taken together, these parameters define a feasibility envelope rather than a calibrated
model. Their purpose is to make implicit governance and implementation constraints explicit
and auditable, not to maximize expected returns. Consequently, different parameter choices
will lead to different admissible satellite designs, but the qualitative structure of the
feasibility hierarchy—and the resulting separation between feasibility and forecasting—
remains unchanged.


\end{document}